\documentclass[11pt,a4paper]{article}

\usepackage{comment}
\usepackage{graphicx}
\usepackage{amsmath,amssymb}
\usepackage{bbm}
\usepackage{hyperref}
\DeclareGraphicsExtensions{.eps}

\begin{document}

\title{Geographical Load Balancing \\ across Green Datacenters: a Mean Field Analysis\footnote{
	Published in the proceedings of Greenmetics workshop 2016, in conjunction with ACM Sigmetrics/IFIP Performance, Juan-les-Pins, France, June 14, 2016
}}
\author{
Giovanni Neglia\\
  Universit\'e C\^ote d'Azur - Inria, France\\ 
  \texttt{giovanni.neglia@inria.fr}
  \and
  Matteo Sereno\\
  Dip. di Informatica,\\
   Universit\`{a} di Torino, Italy\\
  \texttt{matteo.sereno@di.unito.it}
\and
Giuseppe Bianchi\\
  Dip. di Ingegneria Elettronica, \\
  Universit\`{a} di Roma ``Tor Vergata", Italy\\
  \texttt{giuseppe.bianchi@uniroma2.it}
}

\date{June 2016}

\maketitle
\begin{abstract}
``Geographic Load Balancing'' is a strategy for reducing the energy cost of data centers spreading across different terrestrial locations. In this paper, we focus on load balancing among micro-datacenters powered by renewable energy sources. We model via a Markov Chain the problem of scheduling jobs by prioritizing datacenters where renewable energy is currently available. Not finding a convenient closed form solution for the resulting chain, we use mean field techniques to derive an asymptotic approximate model which instead is shown to have an extremely simple and intuitive steady state solution. After proving, using both theoretical and discrete event simulation results, that the system performance converges to the asymptotic model for an increasing number of datacenters, we exploit the simple closed form model's solution  to investigate relationships and trade-offs among the various system parameters.
\end{abstract}
%\keywords{Energy Efficiency, Data Centers, Geographical load balancing, Fluid Models.}

\section{Introduction}
Providers such as Amazon, Google, Facebook, etc., are making a considerable effort to offer efficient, scalable, and reliable services. To achieve these goals such services need to be supported by massive datacenters %(in terms of servers, storage, networking, etc.)
and relevant infrastructures to distribute power and provide cooling. Power management is becoming a crucial issue. Not only power consumption is ever increasing with an increasing user base and service expansion, but, as pointed out by several studies, the power consumption of datacenters is largely wasted.
%and calls for datacenter energy optimization techniques
%\cite{NRDC2015}.

In this paper we consider a set of micro-datacenters which are additionally powered by renewable energy sources, e.g., photovoltaic (PV) panels. Due to the current high costs for storing energy, the best use of renewable energy is to consume it when it is produced. % (or to resell it to the energy provider, but we do not consider this possibility here).
Hence, we would ideally wish to adapt each micro-datacenter's load to the instantaneous energy production.
%To this purpose each micro-datacenter can try to adapt its energy demand for computation and cooling to the instantaneous energy production.
%Nevertheless, the job profile of a single micro-datacenter may not exhibit enough flexibility to follow the high variability of renewable energy production.
One way to address such goal is to {\em federate} several micro-datacenters each other, and use a central controller to dispatch jobs where renewable energy is available, so as to minimize the (non-renewable) energy cost. The possibility to manage more jobs obviously offers a higher flexibility. The law of large numbers guarantees indeed that the aggregated load will be more regular and then easier to exploit for smart  load scheduling as it is the case in a big datacenter. But when local renewable sources are available, a micro-datacenters' federation offers an additional advantage in comparison to a large datacenter: renewable energy production at different locations can be loosely correlated
%(the less correlated the farther the locations are)
and then the aggregated energy production exhibits less variability.

Consider the following ideal case: a set of $N$ identical datacenters, each with independent job arrival processes with rate $\lambda$ and a single server with computing rate $\mu$, and PV panels able to feed the datacenter a fraction $s$ of the time. Compare it with a single datacenter which aggregates locally all the jobs as well as the computing and energy production infrastructure. The total normalized load for the federation of datacenter is $\rho= (N \lambda)/(N \mu)=\lambda/\mu$ with a normalized variability (standard deviation of the number of working servers divided by the average number of working server) equal to $\sqrt{(1-\rho)/(\rho N)}$. Similarly the federation can power through renewables a fraction $s$ of its computing resources with a normalized variability equal to $\sqrt{(1-s)/(s N)}$, if the amounts of renewable energy produced at different datacenters can be considered independent.
The single datacenter manages the same aggregate load with the same normalized variability, but the situation is different energywise.
The single datacenter can be powered by renewables a fraction $s$ of the time, but now the normalized variability is $\sqrt{(1-s)/s}$, if, as it is reasonable to assume in first approximation, all the PV panels at a given location produce (/do not produce) at the same time. 

The example above is clearly over-simplified, it ignores the costs of job dispatchment among the micro-datacenters, the effect of fixed energy costs that are easier to optimize at a single datacenter, the possibility that renewable energy dynamics are too fast to be exploited by smart scheduling strategies, how revenues should be split among the datacenters, etc. Nevertheless, this example highlights a potential benefit from federating micro-datacenters, that is interesting to quantify. As we are going to show below, even simple models for job traffic and energy production lead soon to scenarios for which it is difficult to provide closed-form expressions for the energy cost of a federation of micro-datacenters. One may then need to rely on expensive simulations that hide the role played by the different parameters. For this reason, in this paper we propose a mean field (fluid) model that is asymptotically correct and allows us to derive simple formulas for the main performance metrics, like the expected energy costs of the system.

The paper is organized as follows. After a brief discussion of related work, we introduce the system model in section III, and we provide and justify with both theoretical and simulation results a mean field approximation in section IV. In section V, we exploit the resulting simple model to quantify performance and trade-offs emerging in scenarios characterized by variable renewable energy production across micro datacenters.

\section{Related Work\label{RELATED}}
In Geographic Load Balancing (GLB) systems user requests are initially accepted
by front-end elements and then redirected by a scheduler to geographically distributed datacenters for processing.
The scheduler's decisions may depend on several mutually interacting (and in some case conflicting) objectives such
as minimizing the electricity cost, the carbon-footprint and the response time.
%, reducing transmission delays and transmission
%costs etc.
The paper \cite{Qureshi2009} is one of the first studies about GLB. In particular,
it focusses the attention on the key issues fostering the use of GLB such
as different energy markets (e.g., day-ahead and real-time markets), and temporal or geographical energy price variations.
%, and
%services supported by multiple geographical datacenters that already include
%geographical load balancing mechanisms.
The GLB represents the combination of these basic ingredients with the use of energy related metrics in the scheduler decisions.
In this manner it is possible to account for different workload conditions, and
time and geographical variability of the electricity costs.

In the last years other studies addressed the same problem by adding different scheduling constraints
and/or by optimizing different metrics (see \cite{Rao2010}, \cite{Rao2012}, \cite{Leana2012}, \cite{GuoBatterie2011},
\cite{WiermanOnline2012}, \cite{WiermanSigmetics2011}).
For instance, the papers \cite{Rao2010} and \cite{WiermanSigmetics2011} introduce additional constraints for accounting QoS guarantees;
while the interaction between GLB and smart grids, and then the exploitation of the workload demand-response capability
have been addressed in \cite{Rao2010}.
Furthermore, the interaction of energy storage systems and GLB has been addressed in \cite{GuoBatterie2011}.
Indeed, storage systems can be used to smooth the variability of power supply and this is very important when the datacenters
are powered by renewable sources.

Several studies pointed out that large datacenters are extremely expensive to maintain and this has
encouraged the development of architectures \hspace{1pt} that interconnects multiple micro-datacenters \cite{Bianchini2012}.
This trend influenced our work because workload scheduling among a large number of interconnected datacenters gives rise to
computational problems (e.g., see the summary of the techniques used in geographical load balancing in \cite{RahmanLK14}).

% ----
The works closest to ours are \cite{WiermanSigmetics2011} and \cite{WiermanOnline2012}, where geographical load balancing is driven by time-varying energy prices, that can be due to a significant local production from renewable sources. While in these papers energy prices are considered to be known in advance over some future time-horizon,  in our case renewable energy production is a stochastic process and scheduling is decided on the basis of the current state of the system.

\section{Problem}
\label{s:problem}
We consider a federation of $N$ identical micro datacenters.
The aggregated job arrival process at the federation is modeled as a Poisson process with rate $N \lambda$. The service time of each job is assumed to be exponential with expected value $1/\mu$.\footnote{While we need an underlying Markovian process to correctly derive the asymptotic fluid model, empirical results show that the fluid model does not heavily depend on many of these assumptions. %(see for instance %. For example~
%\cite{mitzenmacher1991phd}).
%shows that the fluid model for the supermarket model is more accurate when service times are deterministic.
}
Each datacenter is connected to the grid but it can be powered also by some renewable source. We consider here that the renewable source can be in two states: in state $\mathcal S$ (sunny) the energy produced by the source is able to power the whole datacenter, in state $\mathcal C$ (cloudy) the energy produced is negligible. Renewable states evolve according to a continuous time Markov Chain. Let $\nu_C$ and $\nu_S$ denote respectively the transition rates from $\mathcal S$ to $\mathcal C$ and from $\mathcal C$ to $\mathcal S$. The model for the renewable source can be made arbitrarily more realistic by adding multiple states.
 For the moment we assume that the  Markov chains associated to renewable sources at different datacenters evolve independently.

When a new job arrives the scheduler dispatches it  i) to a datacenter that is available to process it and in state $\mathcal S$ (i.e.,~currently powered by renewables) if any, otherwise ii) to an available datacenter if any, and as last option iii) to a central waiting queue from which the job will be moved to the first available datacenter.
The system is then operating as an $M/M/N$ queue with the characteristic that available servers in state $\mathcal S$ get jobs with strict higher priority than other servers. Among the work conserving disciplines this intuitively minimizes the total expected energy cost. % (but we do not prove it formally).

The system can be described as a continuous time Markov Chain with state
$(J^N(t),S^N(t),B_S^N(t))$, where $J^N(t)$ is the number of jobs in the system,
$S^N(t)$ is the number of servers in state $\mathcal S$, and
$B_S^N(t)$ is the number of servers busy (i.e., serving a job) and in state $\mathcal S$, all  at time $t$. The Markov chain has a very particular structure: for example $J^N(t)$ itself, representing the number of jobs in a $M/M/N$ queue, evolves as a Markov chain. $S^N(t)$ is  described by a simple Markov chain too. In particular the stationary distributions of $J^N(t)$ and $S^N(t)$ can be derived easily in closed-form. Despite these properties, it is not easy to characterize $B_S^N(t)$ and in particular we have not been able to derive in closed-form its stationary distribution. This is less surprising if we think about a similar problem for parallel queues where the simple join-the-shortest-queue policy couples the status of the different queues so that their stationary distribution can be expressed only as an infinite mixture of geometric distributions~\cite{adan1991}
(there are many works on priority queues and/or shortest queue policies, see for instance \cite{Foley-McDonald2001}
\cite{Harchol-Balter2005}).
Similarly, here our dispatching policy couples the two different states of a server (being busy and being powered by renewables) in a non-trivial way so that it is difficult to characterize the process $B_S^N(t)$, as we need to quantify the energetic savings coming from the federation.

In order to study the system $(J^N(t),S^N(t),B_S^N(t))$ we could resort to simulations or to a numerical solution of the Markov Chain. In both cases the computational cost increases with the number of datacenters $N$.
These difficulties are aggravated if more realistic and then more complex models for traffic arrival process or renewable energy evolution are considered with a potential explosion of the state space. Moreover, the effect of the different parameters can be more difficult to unveil using numerical methods.
For these reasons, as it has been successfully done in other fields, we derive the fluid limit of the Markov chain of interest, that allow us to obtain simple closed-form expressions for the main performance metrics independently from the system size $N$. % The next section describes how to derive the fluid limit.

\section{Fluid Model}
\label{s:fluid}
In this section we show that the stochastic dynamics of the Markov chains $(J^N(t),S^N(t),B_S^N(t))$ converge in probability to a deterministic process as $N$ diverges.\footnote{In what follows, convergence of random variables is always ``in probability.'' We omit to repeat it at each time.}
 More precisely, we will show that if $1/N(J^N(0),S^N(0),B_S^N(0))$ converges to the constant values $(j_0,s_0,b_{s,0})$ when $N$ diverges, then there exists a vector of deterministic functions $(j(t),s(t),b_s(t))$ such that\break
 $(j(0),s(0),b_s(0))=(j_0,s_0,b_{s,0})$ and for any $T>0$:
\[\sup_{0\le t \le T} \left|\left| \frac{1}{N}(J^N(t),S^N(t),B_S^N(t)) - (j(t),s(t),b_s(t))\right|\right|  \underset{P} \rightarrow 0,\]
i.e., the rescaled process converges to $(j(t),s(t),b_s(t))$.

This kind of convergence results has become popular since the seminal work of Kurtz (see for example \cite{kurtz1976}), that shows that the limiting process can be described by a system of differential equations: $d\mathbf{x}/dt=\mathbf{f}(\mathbf{x}(t))$,
%\[\frac{d\mathbf{x}}{dt}=\mathbf{f}(\mathbf{x}(t)),\]
where $\mathbf f()$ is called the limiting drift function. Classic results require $\mathbf f()$ to be a Lipschitz function. By carrying out the usual derivation of the fluid limit for the process $(J^N(t),S^N(t),B_S^N(t))$, the corresponding function $\mathbf f()$ will appear to be discontinuous and then it has not the Lipschitz property. Nevertheless, we can apply more recent and general results from~\cite{gast2012} to show that the dynamics converge to the solution of a system of differential inclusions, i.e. where the function $\mathbf f()$ is replaced by a set valued function.

As we observed in the previous section, the processes $J^N(t)$ and $S^N(t)$ are themselves Markov chains. Rather than studying the joint system\break $(J^N(t),S^N(t),B_S^N(t))$ we first derive the fluid limits for $J^N(t)$ and $S^N(t)$ and then move to consider the fluid limit for $B_S^N(t)$. While we could directly consider the limit of the triplet, this approach can result easier to follow for the reader unfamiliar with fluid limits. Moreover, the results for $J^N(t)$ and $S^N(t)$ do not require the more complex machinery of differential inclusions, so this approach allows us to better highlight where difficulties arise for $B_S^N(t)$.

The Markov chain describing $J^N(t)$ is such that the transition from state $J$ to state $J+1$ occurs with rate $N \lambda$, while the transition from state $J$ to state $J-1$ occurs with rate $\mu J$, if $J \le N$, and with rate $\mu N$, if $J > N$.
We consider now the scaled process $J^N(t)/N$, whose transition rates from $x$ to $x+l/n$ can be expressed as $N \beta_l(x)$ where the functions $\beta_l(x)$ do not depend on $N$. In particular $\beta_1(x)=\lambda$, $\beta_{-1}(x)=\mu x$ for $x\le1$, $\beta_{-1}(x)=\mu $ for $x>1$ and $\beta_l(x)=0$ otherwise.
The rate of changes of $J^N(t)/N$ is then
\[f(x)=\frac{1}{N}\left(\beta_1(x) -\beta_{-1}(x) \right)=\begin{cases}
	\lambda - \mu x & \mbox{if } x\le 1\\
	\lambda - \mu  & \mbox{if } x> 1
\end{cases}
\]
that is a Lipschitz function. This property and the fact that $\sum_l |l| \beta_l(x) < \infty$ guarantee~\cite{kurtz1976} that if $J^N(0)/N$ converges to $j_0$, $J^N(t)/N$ converges to the unique solution\footnote{
	Continuity of the right hand side guarantees the existence of the solution and Lipschitz property guarantees uniqueness.
} of the following equation
\begin{equation}
\label{e:jobs}
\frac{d j}{d t}=f(j(t)), \;\; j(0)=j_0.
\end{equation}
Observe that $j(t)>1$ corresponds to $J^N(t)>N$ and then a situation where all the $N$ data centers are working and there are $J^N(t)-N$ jobs in the queue.
Given that $f(x)<0$ for $x>1$,
Eq.~\eqref{e:jobs} shows that $j(t)<1$ for large enough $t$ and then after
some transient the job queue is asymptotically empty and the number of
jobs in the system coincides with the number of busy servers.
Moreover, $j(t)$ converges when $t$ diverges:
$j^* \triangleq j(\infty)=\lambda/\mu=\rho$.
This value is the only accumulation point for the possible trajectories
of $j(t)$ and then it is also the stationary probability that a server
is busy in the original Markov chain~\cite{benaim2008}
(as it is known from the analysis of the $M/M/N$ queue).

In a similar way, it is possible to show that if $S^N(0)/N$ converges  to $s_0$, $S^N(t)/N$ converges to the solution of the following equation
\begin{equation}
\label{e:suns}
\frac{d s}{d t}=\nu_S-(\nu_S+\nu_C) s(t), \;\; s(0)=s_0,
\end{equation}
and when $t$ diverges $s(t)$ converges to $s^* \triangleq \nu_S/(\nu_S+\nu_C)$, that is the stationary probability that a given datacenter is powered by renewables.

It is clear that we would not have needed fluid models to derive the asymptotic probability that a datacenter is busy or that it is powered by the renewables, but the fluid models allow us to evaluate simply the transient dynamics for the percentage of busy datacenters and of datacenters powered by renewables. Moreover, they are required to characterize the quantity $B_S(t)/N$ that is needed to quantify how many datacenters work using the cheap renewable energy.

\begin{figure}
\centering
\includegraphics[width=0.7\textwidth]{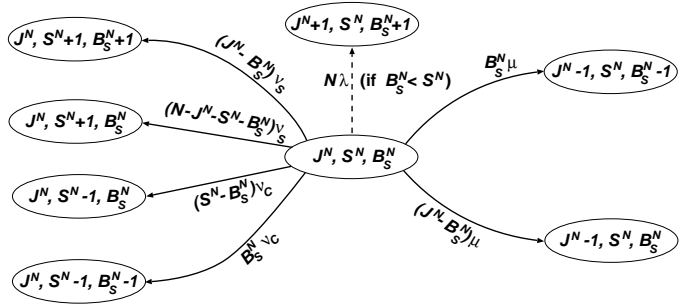}
\caption{Transitions that bring to a change in $B_S$ for $J\le N$.}
\label{f:markov_chain}
\end{figure}

In Fig.~\ref{f:markov_chain} we show the Markov chain transitions affecting $B_S(t)$, i.e.~the number of datacenters working and powered by renewables, when the number of jobs in the system $J(t)$ is smaller than $N$.  As we observed above, for $N$ large enough $J(t)<N$ holds with probability arbitrarily close to one after some finite time depending on $J^N(0)$. For this reason, we can for simplicity assume that the system is in this situation. Observe that the transition indicated in the figure by the dashed line is possible only for specific values of $S$ and $B_S$. If a new job arrives and there are idle datacenters in state $\mathcal S$ (i.e. $B_S<S$) then the job will be assigned to one of them and $B_S$ will increase by one unit. Otherwise $B_S$ will stay constant.   If we calculate the drift for $B_S(t)/N$ when $(J(t),S(t),B_S(t))=(j,s,b_s)$ as done above we obtain that it is equal to
\[g(j,s,b_s)=\begin{cases}
	\lambda - (\nu_S+\nu_C+\mu) b_s + \nu_S j & \mbox{if  }b_s<s,\\
	- (\nu_S+\nu_C+\mu) b_s + \nu_S j & \mbox{if  } b_s=s.\\
\end{cases}\]
Unfortunately the function $g()$ is not continuous, and then neither Lipschitz.
%Moreover it cannot exist a differentiable function whose derivative has a discontinuity of second kind as $g(j,s,b_s)$ would imply.
Nevertheless, \cite{gast2012} shows that when $B^N_S(0)/N$ converges in probability to $b_{s,0}$, $B^N_S(t)/N$ is related to the solutions of the following differential inclusion
\begin{align}
\label{e:working_suns}
\frac{d b_s}{d t}& =G(j,s,b_s)=\begin{cases}
	\{g(j,s,b_s)\} & \mbox{if  }b_s<s,\\
	[g(j,s,s),g(j,s,s)+\lambda] & \mbox{if  } b_s=s.\\
\end{cases}\\
b_s(0)& =b_{s,0}\nonumber
\end{align}
The set-valued function $G(j,s,b_s)$ coincides with $g(j,s,b_s)$ for $b_s<s$, while $G(j,s,s)$ is the interval obtained by the convexification of the accumulation points of $g(j,s,b_s)$ when $b_s=s$.
Equation~\eqref{e:working_suns} admits at least a solution because $G()$ is upper-semicontinuous and Theorem~5 in \cite{gast2012} shows that in such case
\[\inf_{b_s \in \mathcal D} \sup_{0\le t\le T} \left|\left| \frac{B_S(t)}{N}- b_s(t)\right|\right|\underset{P}\rightarrow 0, \]
where $\mathcal D$ is the set of solutions of Eq.~\eqref{e:working_suns}. This result has practical utility if we can prove that the differential inclusion~\eqref{e:working_suns} has a unique solution. A standard sufficient condition for the uniqueness of the solution is the one side Lipschitz condition~\cite{kunze2000}, that unfortunately does not hold for $G()$. We suspect that Eq.~\eqref{e:working_suns} has a unique solution, but we have not been able to prove it. Nevertheless, we can prove that any possible solution converges to the same value as $t$ diverges. This is enough to draw conclusions about the stationary distribution of our stochastic system.

We start observing that for large $t$ $j(t)$ and $s(t)$ are arbitrarily close respectively to the values $j^*=\rho$ and $s^*$. It holds
\[g(j^*,s^*,s^*)= \nu_S \left( \rho -\frac{\nu_S+\nu_C+\mu}{\nu_S+\nu_C}   \right)  <0,\]
because $\rho<1$. If $\lambda+g(j^*,s^*,s^*)<0$, then all the values of $G(j^*,s^*,b_s)$ are negative when $b_s$ belongs to an opportune interval $(s^*-\epsilon, s^*]$  and then any possible trajectory of $b_s(t)$ will be constrained to the interval $[0,s^*-\epsilon]$, where the differential inclusion~\eqref{e:working_suns} reduces to a usual differential equation with Lipschitz drift and then it admits a unique solution. This solution converges to $(\lambda +\rho \nu_S)/(\nu_S+\nu_C+\mu)$ when $t$ diverges.
If $\lambda+g(j^*,s^*,s^*)>0$, then for $b_s < s^* $
\[g(j^*,s^*,b_s)>\lambda -(\nu_S+\nu_C+\mu) s^* + \nu_S j^* =\lambda+g(j^*,s^*,s^*)>0,\]
and any trajectory of $b_S(t)$ converges to $s^*$, that is a stable point because in this case $0 \in G(j^*,s^*,s^*)$.
Summarizing, it holds
\[b_s^*\triangleq b_s(\infty)=\begin{cases}
	s^*, \mbox{   if } \lambda+g(j^*,s^*,s^*)>0\\
	(\lambda +\rho \nu_S)/(\nu_S+\nu_C+\mu), \mbox{ otherwise.}
\end{cases}
\]
By observing that $\lambda+g(j^*,s^*,s^*)>0$ is equivalent to $s^* < (\lambda +\rho \nu_S)/(\nu_S+\nu_C+\mu)$, and replacing $\lambda=\rho \mu$ we can write in a more compact way:
\begin{equation}
	\label{e:ws_asymp}
	b_s^*= \min\left\{s^*, \rho \frac{ \nu_S + \mu}{\nu_S+\nu_C+\mu}\right\}.
\end{equation}
%This equation shows that there are two regimes in the system. In one the percentage of

\begin{figure}[bht]
\centering
\includegraphics[angle=270, width=0.7\textwidth]{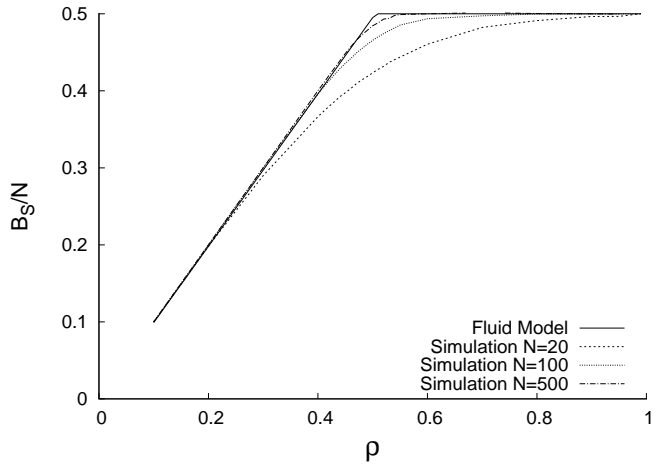}
\caption{Percentage of datacenters powered by renewables
: Fluid Model ($b_s^*$) vs Simulation Averages
($\nu_S=\nu_C=0.01$ $\mu=1$, $N=20, 100, 500$).}
\label{f:validation}
\end{figure}

Figure~\ref{f:validation} shows how the stationary distribution of $B_S^N/N$ converges to $b_s^*$ as $N$ increases. The quality of the fluid approximation is different for   different values of the load $\rho$.
In particular as far as $\rho$ is far from the critical value for which $s^*= \rho (\nu_S + \mu)/(\nu_S+\nu_C+\mu) $, corresponding to the non differentiability in Eq.~\eqref{e:ws_asymp}, the approximation is very accurate even for $N=20$ datacenters. For the critical load, the federation should include an order of magnitude more datacenters to achieve a good level of approximation.
For a given value of $N$ the quality of the approximation improves (/worsen) the larger (/smaller) is the acute angle between the two segments determined by the fluid model, as it happens if $\nu_C$ increases (/decreases).

%If the right hand side is negative, the drift will always be negative and the trajectory moves away to
%
%When $b_s$ is arbitrarily close to $s^*$, the drift is $g(j^*,s^*,s^{*-})=\lambda+g(j^*,s^*,s^*)$. If $g(j^*,s^*,s^{*-})$ is also negative, then the trajectory

%In order to derive the fluid limit for $J^N(t)/N$, it is useful to consider an equivalent discrete time model with time slot (proportional to) $1/N^2$: $j^N_k=J^N(k/N^2)/N$.\footnote{
%	This can be made formally rigorous considering the uniformization of the Markov chain, see \cite[Appendix B.3]{gast2012}
%}. After one slot the rescaled process $j^N_k$ moves from state $x$ to state $x+1/N$ with probability $\lambda/N$ and to state  $x-1/N$ with probability $\mu x/N$ if $x\le1$ and with probability $\mu$ if $x >1$. The expected increment of the process $J^N/N$ in $x$ (called the drift) is then:
%\[\textrm{E}\left[j^N_{k+1}-j^N_k | j^N_k=x \right]=\begin{cases}
%	\frac{1}{N}\left(\lambda - \mu x \right)& \mbox{if } x\le 1\\
%	\frac{1}{N}\left(\lambda - \mu \right) & \mbox{if } x> 1
%\end{cases}
%\]
%The drift goes to zero with rate $1/N$, by normalizing we can obtain the limiting drift
%\[f(x)=\begin{cases}
%	\lambda - \mu x & \mbox{if } x\le 1\\
%	\lambda - \mu  & \mbox{if } x> 1
%\end{cases}
%\]
%that is a Lipschitz function. This property together with the fact that the $J(t)$ can increase/decrease only by one unit at any time\footnote{
%	It would be enough to show that the sum over all the possible
%}

\section{Exploiting the model}
In this section we show how our simple fluid model can help quantifying the potential advantages of a federation of datacenters and the effect of the different parameters.

We start by discussing Eq.~\eqref{e:ws_asymp}.
The percentage $b^*_s$ of datacenters working and powered by renewables is obviously limited by the percentage $s^*$ of datacenters powered by renewables, and by the percentage $\rho$ of datacenters working, then $b^*_s \le \min\{s^*,\rho\}$.
These two regimes appear also in Eq.~\eqref{e:ws_asymp} and we refer to them as the renewables-limited regime and the load-limited regime.
In particular, Eq.~\eqref{e:ws_asymp} shows how close  the dispatching algorithm can approach the bound $\min\{s^*,\rho\}$ when, as we assumed, the job will be completed by the datacenter that started working on it. The factor $(\nu_S+\mu)/(\nu_S+\nu_C+\mu)$ multiplying $\rho$ takes into account the fact that a datacenter may change status from $\mathcal S$ to $\mathcal C$ (or the other way around) after starting to process a job. These changes limit the utility of job scheduling.

Without the federation every datacenter receives a load $\rho$ and can exploit renewables a fraction $s^*$ of the time. Then the percentage of time a datacenter works and is powered by renewables is $\rho s^*$, that is smaller than $b^*_s$ from Eq.~\eqref{e:ws_asymp}:
\[\min\left\{ s^*, \rho \frac{\nu_S+\mu}{\nu_S+\nu_C+\mu}\right\}> \rho s^*,\]
because $\rho <1$ and $(\nu_S+\mu)/(\nu_S+\nu_C+\mu) > \nu_S/(\nu_S+\nu_C)=s^*$.
The difference between the left hand side and the right hand side of the inequality times $N$ quantifies how many additional datacenters work powered by renewables thanks to the federation in comparison to the situation when there is no federation.
In what follows we compare the corresponding average energy costs, by normalizing the energy cost per time unit to $1$ (/$0$) when the datacenter is (/is not) powered by renewables. The average energy cost per time unit and per datacenter is then:
\begin{align*}
c_f &  \triangleq  \rho - b^*_s,  \mbox{  with the federation}\\
c_{nf} & \triangleq  \rho - \rho  s^*,  \mbox{without the federation}
\end{align*}

\begin{figure}[bht]
\centering
\includegraphics[angle=270, width=0.7\textwidth]{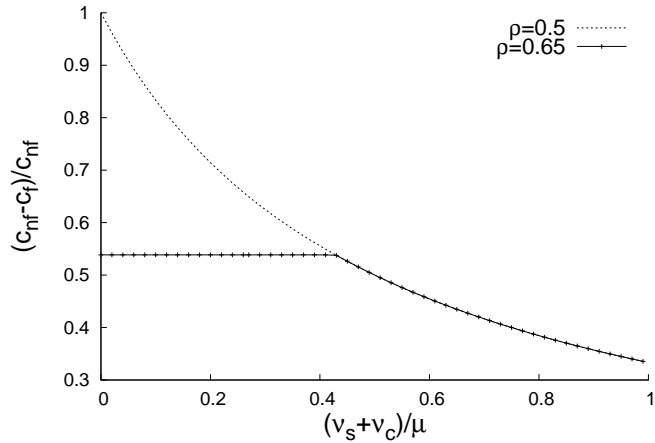}
\caption{Cost reduction due to the federation vs speed of renewables' dynamics ($s^*=0.5$).}
\label{f:dynamics_ratio}
\end{figure}

We focus on the relative cost reduction achieved by the federation in comparison to the uncoordinated case, i.e., on $(c_{nf}-c_f)/c_{nf}$.
Fig.~\ref{f:dynamics_ratio} shows how the relative cost changes as renewables' dynamics become faster for two different values of the load $\rho$. We set $\nu_S=\nu_C$ so that $s^*$ is constant and equal to $0.5$. Eq.~\eqref{e:ws_asymp} shows that, when $\rho$ and $s^*$ are constant, $b^*_s$ changes only for the effect of the ratio $(\nu_S+\nu_C)/\mu$. In other words, it is not important how fast the quantity of renewable energy produced changes, but how much faster it changes than the job completion time. Intuitively, if this ratio is very large, the scheduling is not effective, because a datacenter changes its status $\mathcal S$/$\mathcal C$ many times before completing the job, so that the job takes advantage of renewables' energy on average a fraction $s^*$ of the time, independently from the status of the datacenter when the job execution started. Fig.~\ref{f:dynamics_ratio} shows indeed that the advantage of the federation converges to $0$ as the ratio $(\nu_S+\nu_C)/\mu$ diverges.
This behaviour is common to both the load values considered. When $\rho=0.5$, the system is always in the load-limited regime ($b^*_s=\rho (\nu_S+\mu)/(\nu_S+\nu_C+\mu)$) and the advantage of the federation always decreases as the ratio $(\nu_S+\nu_C)/\mu$ increases. When $\rho=0.65$ the system is initially in the renewables-limited regime, so that the relative gain of the federation is limited by the average availability of renewables' energy and the gain is independent on the speed of their dynamics. This situation corresponds to the initial horizontal part of the corresponding curve. As the speed of renewables' dynamics further increases, the scheduling is no more able to effectively follow them and the system enters in the load-limited regime.
The relative improvement from the federation in this regime is independent from $\rho$, so that both curves in Fig.~\ref{f:dynamics_ratio} overlap.

Our analysis shows significant reduction of energy costs achievable by the federation of different datacenters, but, until now, we have assumed that the states of the renewables' sources at the different datacenters are independent. This is not true in general.  For example, production from PV panels or wind turbines are clearly positively correlated at nearby locations. When energy quantities produced at the datacenters are positively correlated, the improvement from scheduling is reduced. In order to quantify the effect of positive correlation, we consider the following simple model. We assume that the Markov chain determining the state of a renewable source ($\mathcal S$ or $\mathcal C$) is modulated by an underlying Markov chain that is common to all the different sources.
In particular, as a toy-example, we consider a Markov chain with two states $\mathcal G$ and $\mathcal B$. The transition rates $\nu_S$ and $\nu_C$ of each renewable source depend now on the particular state of the modulating Markov chain and we denote them $\nu_{S,G}$,  $\nu_{C,G}$, $\nu_{S,B}$ and $\nu_{C,B}$.
We consider that
\[s^*_G=\frac{\nu_{S,G}}{\nu_{S,G}+\nu_{C,G}}>\frac{\nu_{S,B}}{\nu_{S,B}+\nu_{C,B}}= s^*_B\]
and then states $\mathcal G$ and $\mathcal B$ correspond respectively to good and bad weather (at least for the purpose of renewable energy production).
It is possible to extend simply our previous analysis, if we assume that the dynamics of the modulating Markov chain are much slower than those of the modulated chain and of  job execution (i.e., $\max\{\nu_G, \nu_B\} <\!< \min\{\nu_{S,G}, \nu_{C,G},\nu_{S,B}, \nu_{C,B}, \mu\} $). In such case, the average percentage of datacenters working and powered by datacenters can be obtained through a weighted sum of what would happen without modulation as follows
\[b_s^* \approx \pi_G b_{s,G}^*+ \pi_B b_{s,B}^*,\]
where $\pi_G= \nu_G/(\nu_G+\nu_B)$, $\pi_B= \nu_B/(\nu_G+\nu_B)$ and  $b_{s,G}^*$ (resp. $b_{s,B}^*$) is calculated from Eq.~\eqref{e:ws_asymp} replacing the rates $\nu_S$ and $\nu_C$ by $\nu_{S,G}$ and $\nu_{C,G}$ (resp. $\nu_{S,B}$ and $\nu_{C,B}$).
As we anticipated, the modulating Markov chain correlates the state of the renewable sources. 
We can quantify 
this effect by using the correlation coefficient 
% defined as % of the two indicator random variables for datacenters $i$ and $j$ with $i\neq j$: 
% $\mathbb{1}(\mbox{$i$ is in state $\mathcal S$})$, 
% $\mathbb{1}(\mbox{$j$ is in state $\mathcal S$})$. 
%This correlation coefficient 
$\eta$ defined as % is equal to
\[\eta=\frac{(s^*_G)^2 \pi_G+(s^*_B)^2 \pi_B - (s^*_G \pi_G+s^*_B \pi_B)^2}{(s^*_G \pi_G+s^*_B \pi_B)(1- s^*_G \pi_G -s^*_B \pi_B)}.\]
As a sanity check, we observe that if $\nu_{S,G}=\nu_{S,B}$ and $\nu_{C,G}=\nu_{C,B}$ (i.e., the modulating Markov Chain has no effect on the renewables' state evolution), then $\eta=0$. If instead we have that $s^*_G=1$ and $s^*_B=0$, then $\eta=1$, because all the datacenters are in state $\mathcal S$ when the modulating Markov chain is in state $\mathcal G$ and in state $\mathcal C$ when the modulating Markov chain is in state $\mathcal B$.

\begin{figure}[bht]
\centering
\includegraphics[angle=270, width=0.7\textwidth]{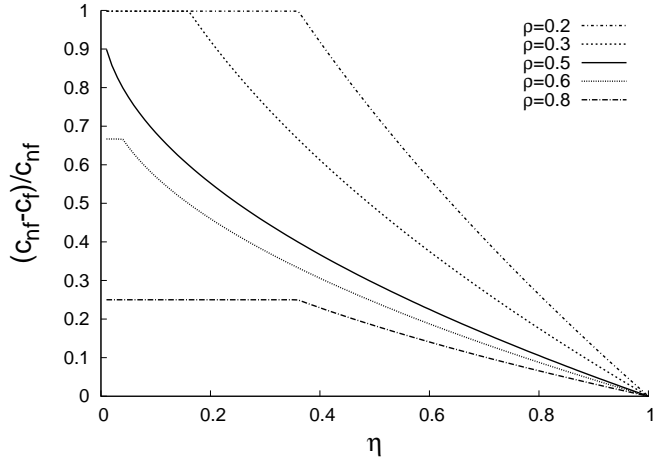}
\caption{Cost reduction due to the federation vs renewables' correlation ($\nu_G=\nu_B=0.00001$, $\nu_{S,G}+\nu_{C,G}=\nu_{S,B}+\nu_{C,B}=0.002$, $\nu_{S,G}=\nu_{C,B}$, $\mu=1$).}
\label{f:correlation}
\end{figure}

Figure~\ref{f:correlation} shows the relative cost reduction due to the federation versus the correlation $\eta$.
In the specific setting considered, the average percentage of time renewables can power datacenters is constant: $s^*=\pi_G s^*_G+\pi_B s^*_B=0.5$. Then, as the correlation increases $s^*_G$ increases and $s^*_B$ decreases of the same amount.
As expected, the benefit from the federation is maximum when renewable sources evolve independently ($\eta=0$) and null when at any time they are all in the same state ($\eta=1$). The benefit is non-increasing in $\eta$, but, depending on the load $\rho$, there is a more or less wide range of correlation values for which the benefit does not depend on $\eta$.
In order to justify this result, we write the specific expression of  $b^*_s$ neglecting for simplicity the rates $\nu_{S,G}$, $\nu_{C,G}$, $\nu_{S,B}$ and $\nu_{C,B}$ when summed to $\mu$, that is much larger. It holds:
\[b^*_s\approx\frac{1}{2}\min\left\{s^*_G, \rho\right\}+\frac{1}{2}\min\left\{s^*_B, \rho\right\}\]
Under this approximation, the setting $\rho=0.5=s^*$ corresponds to the case when the system is at the boundaries between the two regimes for $\eta=0$. When the correlation increases, the system is i) in the load-limited regime in good weather (state $\mathcal G$) with a value $b^*_{s,G}$ almost constant and equal to $\rho$ and ii) in the renewables-limited regime in bad weather (state $\mathcal B$) with a value $b^*_{s,B}$ decreasing in $\eta$. As a consequence the corresponding curve is decreasing.
When $\rho>s^*=0.5$, the system is the renewables-limited regime in both states $\mathcal G$ and $\mathcal B$ when $\eta=0$, and then $b^*_{s,G}=s^*_G$ and $b^*_{s,B}=s^*_B$. As $\eta$ increases, the increase of $s^*_G$ is exactly compensated by the decrease of $s^*_B$ so that the system exhibits the same relative improvement until $\eta$ is so large that the system enters in the renewables-limited regime when in bad weather and then the improvement decreases again.
Finally, when $\rho<s^*=0.5$, the system is initially in the load-limited regime in both states, and then $b^*_{s,G}=b^*_{s,B}=\rho$, independent on $\eta$. Again, the improvement does not depend on $\eta$ until $\eta$ becomes so large that the system enters in the renewables-limited regime when in bad weather.

As we have shown, our simple fluid model reveals the existence of two different regimes and helps to understand and quantify their non-trivial interaction as the parameters change.

\section{Conclusions}
The paper proposes a model of geographical load balancing strategies for a collection 
of federated (micro) datacenters powered by renewable energy sources.
In our strategy the scheduler uses a selection criterion that prioritizes 
datacenters where renewable energy is currently produced. 
For this kind of system we use mean field techniques to derive a simple approximate model that allows us to derive several performance measures.
First, asymptotic convergence is proven and the quality of the approximation for finite size systems is evaluated through an ad-hoc simulator.
Then, we use the simple fluid model to quantify the effect of the different system parameters and to understand the different tradeoffs.

\section*{Acknowledgements}
This work was supported in part by the "Investments for the Future"
Program reference
\#ANR-11-LABX-0031-01, funded by the French Government
(National Research Agency, ANR), and in part by the
Universit\`{a} Italo-Francese, call Galileo 2015-2016, ref. G15-133.

\bibliographystyle{abbrv}

\bibliography{REF}

\begin{thebibliography}{10}

\bibitem{adan1991}
I.~J. B.~F. Adan, J.~Wessels, and W.~H.~M. Zijm.
\newblock {Analysis of the Asymmetric Shortest Queue Problem}.
\newblock {\em Queueing Systems}, 8(1):1--58, 1991.

\bibitem{benaim2008}
M.~Bena{\"i}m and J.~Y. Le~Boudec.
\newblock {A Class of Mean Field Interaction Models for Computer and
  Communication Systems}.
\newblock {\em Performance Evaluation}, 65(11-12):823 -- 838, 2008.

\bibitem{Bianchini2012}
R.~Bianchini.
\newblock {Leveraging Renewable Energy in Data Centers: Present and Future}.
\newblock In {\em Proc. of HPDC}, 2012.

\bibitem{Foley-McDonald2001}
R.~D. Foley and D.~R. McDonald.
\newblock {Join the Shortest Queue: Stability and Exact Asymptotics}.
\newblock {\em The Annals of Applied Probability}, 11(3):569--607, 2001.

\bibitem{gast2012}
N.~Gast and B.~Gaujal.
\newblock Markov chains with discontinuous drifts have differential inclusion
  limits.
\newblock {\em Performance Evaluation}, 69(12):623--642, 2012.

\bibitem{GuoBatterie2011}
G.~Guo, Z.~Ding, Y.~Fang, and D.~Wu.
\newblock {Cutting Down the Energy Cost of Geographically Distributed Cloud
  Data Centers by Using Energy Storage}.
\newblock In {\em Proc. of GLOBECOM}, 2011.

\bibitem{Harchol-Balter2005}
M.~Harchol-Balter, T.~Osogami, A.~Scheller-Wolf, and A.~Wierman.
\newblock Multi-server queueing systems with multiple priority classes.
\newblock {\em Queueing Systems}, 51(3):331--360, 2005.

\bibitem{kunze2000}
M.~Kunze.
\newblock {\em Non-Smooth Dynamical Systems}.
\newblock Lecture Notes in Mathematics. Springer, 2000.

\bibitem{kurtz1976}
T.~G. Kurtz.
\newblock {Limit Theorems and Diffusion Approximations for Density Dependent
  Markov Chains}.
\newblock In R.~J.-B. Wets, editor, {\em Stochastic Systems: Modeling,
  Identification and Optimization, I}, pages 67--78. Springer, 1976.

\bibitem{WiermanOnline2012}
M.~Lin, Z.~Liu, A.~Wierman, and A.~L. H.
\newblock {Online algorithms for geographical load balancing}.
\newblock In {\em Proc. of IGCC}, 2012.

\bibitem{WiermanSigmetics2011}
Z.~Liu, M.~Lin, A.~Wierman, S.~Low, and A.~L. H.
\newblock {Greening Geographical Load Balancing}.
\newblock In {\em Proc. of ACM SIGMETRICS}, 2011.

\bibitem{Qureshi2009}
A.~Qureshi, R.~Weber, H.~Balakrishnan, J.~Guttag, and B.~Maggs.
\newblock {Cutting the Electric Bill for Internet-scale Systems}.
\newblock In {\em Proc. of ACM SIGCOMM}, 2009.

\bibitem{RahmanLK14}
A.~Rahman, X.~Liu, and F.~Kong.
\newblock {A Survey on Geographic Load Balancing Based Data Center Power
  Management in the Smart Grid Environment}.
\newblock {\em {IEEE} Communications Surveys and Tutorials}, 16(1):214--233,
  2014.

\bibitem{Rao2010}
L.~Rao, X.~Liu, L.~Xie, and W.~Liu.
\newblock {Minimizing Electricity Cost: Optimization of Distributed Internet
  Data Centers in a Multi-Electricity-Market Environment}.
\newblock In {\em Proc. of INFOCOM}, 2010.

\bibitem{Rao2012}
L.~Rao, X.~Liu, L.~Xie, and W.~Liu.
\newblock {Coordinated Energy Cost Management of Distributed Internet Data
  Centers in Smart Grid}.
\newblock {\em IEEE Transactions on Smart Grid}, 3(1):50--58, March 2012.

\bibitem{Leana2012}
Y.~Yao, L.~Huang, A.~Sharma, L.~Golubchik, and M.~Neely.
\newblock {Data centers power reduction: A two time scale approach for delay
  tolerant workloads}.
\newblock In {\em Proc. of INFOCOM}, 2012.

\end{thebibliography}

\end{document}